# Active Flow Control of NACA 0012 airfoil using Sawtooth Direct Current Augmented Dielectric Barrier Discharge Plasma Actuator

Ravi Sankar Vaddi[1], Charles Sota[2], Alexander Mamishev[3], Igor V Novosselov[1,4,*]

[1]Department of Mechanical Engineering, University of Washington, Seattle, U.S.A. 98195

[2]Aerojet Rocketdyne Holdings, Redmond, U.S.A. 98052

[3]Electrical and Computer Science Department, University of Washington, Seattle, U.S.A. 98195

[4]Institute for Nano-Engineered Systems, University of Washington, Seattle, U.S.A. 98195

## ABSTRACT

Dielectric barrier discharge (DBD) plasma actuators are an attractive option for separation control, lift enhancement, and drag reduction. Some plasma actuators feature optimized electrode shapes, electrical waveforms to maximize the aerodynamic forces at high angles of attack. Here, we analyze the performance of a direct current augmented DBD (DBD – DCA) actuator with a sawtooth shape exposed electrode. The active electrode was positioned at 18% chord and the DCA electrode at 48% chord of NACA 0012 airfoil. Wind tunnel experiments were conducted at wind speeds of 15 – 25 m/s, corresponding to Reynolds numbers Re = $2.01 \times 10^5$ – $3.35 \times 10^5$. Lift coefficient ($C_L$), drag coefficient ($C_D$), and pitching moment coefficients ($C_M$), were measured with and without plasma actuation for angles of attack $\alpha$ = 0° – 8° and the DCA electrode potential ($\varphi_{DC}$) was varied from 0 kV to -15 kV. With energized DCA electrode, the $C_L$ increases up to 0.03 and the $C_D$ decreases by 50% at 15 m/s flow speeds and 0° angle of attack, the results are similar throughout the range of α. The effect of the actuator at higher Re diminishes, suggesting that the maximum control authority could be achieved at lower wind speeds.

## 1. INTRODUCTION

Active flow control has been a popular topic in fluid mechanics for the past two decades; the approach showed promise for aerial vehicle maneuvering. It involves small-scale actuators that can create flow field modification at the surface, eliminating the need for traditional control surfaces. Plasma actuators have been studied due to their low profile and simple design while being capable of altering the flow characteristics in the boundary layer without moving parts, with low energy consumption, and high momentum injection (Corke et al. 2010, Corke et al. 2009, Guan et al. 2018). In a corona discharge or dielectric barrier discharge (DBD), the ions are generated when a sufficiently high voltage is applied to the electrodes. The air near the surface of the emitting electrode breaks down and is ionized. The charged species are accelerated in the electric field, exchanging momentum with neutral molecules (Guan, Vaddi, Aliseda and Novosselov 2018, Townsend 1914, Vaddi et al. 2020). The momentum generated by the plasma actuator can be utilized for boundary layer control (Zhang et al. 2019), turbulent mixing layers (Singh and Little 2020), electric propulsion (Hari Prasad et al. 2020, Xu et al. 2018), laminar turbulence transition delay (Dörr and Kloker 2017, Szulga et al. 2015), lift augmentation (Feng et al. 2017, Kotsonis et al. 2014), drag reduction (Kim et al. 2020), and separation control (Messanelli et al. 2019, Wang et al. 2017).

One of the widely used plasma actuators is the alternating current (AC) DBD plasma actuator, which is driven by AC high voltage with frequencies in the order of kilohertz. A DBD actuator comprises two electrodes (one exposed and one encapsulated) separated by an insulating material, and the electrodes are installed in asymmetric arrangement to the dielectric layer. The wall jet generated by the DBD injects momentum into the boundary layer changing the airfoil's aerodynamic characteristics. For example, a DBD actuator installed on the leading edge was shown to suppress flow separation on a NACA $66_3$-018 airfoil, achieving a delay in a stall angle by $8^0$ and increasing the lift to drag ratio by 400 % (Post and Corke 2004). Tsubakino et al. studied the effect of actuator positions 5%, 10%, and 20% from the leading edge to control the flow separation for NACA 0012 airfoil (Tsubakino et al. 2007). The actuators installed as close as possible to the leading edge, upstream of the separation point, effectively control flow separation. For example, Moreau et al. installed three DBD actuators on a NACA 0015; when all the actuators were energized, the separation point moved from 0.5 to 0.76 chord ($c$) (Moreau et al. 2016). A comprehensive review of DBD actuators in the flow control applications can be found in refs (Benard and Moreau 2014, Wang et al. 2013).

---

[*] ivn@uw.edu

The standard DBD configuration for flow control is limited due to low electrical to mechanical energy conversion efficiencies and poor performance at high flow velocities (Tang et al. 2021). Contoured exposed electrode shapes have been shown to increase the actuator performance by introducing a three-dimensional flow field. One of the most effective shapes is a triangular (sawtooth, serrated) electrode. It acts as a vortex generator as the momentum is injected perpendicular to the electrode's edge. The induced vortices increase mixing between the freestream and the boundary layer, energizing the boundary layer (Jukes and Choi 2013). Thomas et al. reported that the sawtooth-type electrodes increased the thrust force up to 50% (Thomas et al. 2009). Messanelli et al. compared the effect of the straight-edge electrode and the sawtooth electrode on lift, drag, and stall angle of NACA 0015 airfoil; they found that sawtooth electrodes are favorably compared to straight electrodes at $Re = 330k$ (Messanelli, Frigerio, Tescaroli and Belan 2019). A similar comparison was performed by Wang for NACA 0015 airfoil at $Re = 77k$; sawtooth actuator led to a stall angle delay by 5° and an increase in the maximum lift coefficient by ~ 9%. At the same time, the traditional DBD plasma actuator delayed stall by only 3°, and a lift increase was ~ 3% (Wang, Wong, Lu, Wu and Zhou 2017).

The DBD induced wall jet's velocity can be further increased by introducing a third exposed electrode creating a sliding discharge plasma actuator (Sosa et al. 2008). Sliding DBD (SDBD) extends the plasma discharge length increasing the body force acting on fluid near the surface (Chen et al. 2020, Zheng et al. 2020). Matsuno et al. experimentally demonstrated that the SDBD thrust increases up to a certain voltage on the third (sliding) electrode and then sharply decreases (Matsuno et al. 2016) due to the formation of a reversed jet from the sliding electrode. To overcome this problem, the SDBD can be further modified where the two electrode DBD ionizes the gas, and the third electrode is used to extend the electrical field in the streamwise direction. The method accelerates the ions and fosters their interaction with neutral molecules over a longer distance; this scheme is called DC augmented DBD (DBD – DCA) (Vaddi et al. 2021, Vaddi et al. 2021). The surface-mounted DBD – DCA actuator with negative DCA potential generates 2x more thrust force than standard DBD (Vaddi, Mamishev and Novosselov 2021). Thrust can be increased by a factor of 4x when the sawtooth electrode is used (Vaddi, Mamishev and Novosselov 2021). Analogous decoupled plasma actuator approaches have been studied for propulsion applications (Xu, He, and Barrett 2019; Gomez-Vega et al. 2021) and have reported a significant increase in thrust and thrust to power ratio.

Most of the studies on flow control over an airfoil used DBD actuators at high angles of attack, where momentum injection was used to overcome the adverse pressure gradient and to trigger the transition to turbulence with DBD actuator acting as vortex generator (VG) at the leading edge of the airfoil. Very few reports have characterized the effect of three-electrode DBD on airfoil performance at lower $\alpha$. This study characterizes the effect of sawtooth DBD – DCA on the performance of NACA 0012 airfoil at low angles of attack in a subsonic wind tunnel. Force measurements were obtained to calculate the lift and drag coefficients and the pitching moment coefficient for $\alpha = 0^0 - 8^0$. The effect of the plasma actuator on the aerodynamic coefficients was determined for three Reynolds numbers. The effects of the DC voltage on the third electrode were also studied to estimate the control efficiency.

## 2. EXPERIMENTAL SETUP AND DIAGNOSTICS

### 2.1. Wind Tunnel and Airfoil

The experiments were conducted in an open return subsonic wind tunnel at the University of Washington with a 0.7 m × 0.383 m rectangular cross-section and a 1.2 m long test section. The tunnel consists of a modular inlet with a series of 10 screens to condition the flow. Inlet is followed by a settling chamber and 10:1 contraction that attaches to the test section. The sidewalls of the test section are plexiglass, allowing for optical access. Downstream of the test section is a short diffuser section connected to a 40 hp, 3 – phase 460 VAC blower controlled by a variable frequency drive. The schematic of the setup is shown in Figure 1.

The NACA 0012 airfoil was used in this study. This airfoil is chosen for its generic shape and well-known aerodynamic characteristics. The active control of the airfoil was also studied extensively, e.g., control of dynamic stall using plasma actuators (Abdelraouf et al. 2020, Abdollahzadeh et al. 2018, Whiting et al. 2020), plasma slats, and flaps (Feng et al. 2012, Feng, Shi and Liu 2017, Zhang et al. 2009). These studies provided data on loads and flow visualization. The airfoil has a 190.5 mm chord ($c$) and 381 mm span ($b$). The size of the airfoil was selected to minimize the blockage effects and while maintaining a large chord Reynolds number. The airfoil was machined from polyurethane foam (Obomodulan) using traditional CNC machining. The endplates, made from the same material as an airfoil, are attached to the top and bottom of the airfoil, flush with walls of the test section, to minimize three-dimensional flow effects. The angle of attack $\alpha$ is set using an optical positioning system (Avago AS 38) with accuracy $\pm 0.02^0$. A stepper motor is used to control the angular position of the airfoil. The angle of attack was varied from $\alpha = 0° - 8°$ during the experiments. Three freestream velocities ($U_\infty$) of 15 m/s (~ 35 mph), 20 m/s (~ 45 mph) and 25 m/s (~ 55 mph) were investigated. These velocities are comparable to those for a typical fixed-wing UAV. The corresponding chord Re numbers are 201k, 268k, and 335k. The $Re$ is defined as

$$Re = \frac{U_\infty c}{\nu} \quad (1)$$

where $\nu$ is the kinematic viscosity. In this study, the maximum angle of attack is 8°, so the blockage ratio corrections are omitted from the $Re$ calculation.

The airfoil is mounted vertically, and the endplates are connected to the external force balance on the top and bottom of the test section. The entire assembly is rested on an air bearing and connected to the bottom load balance flexure. The top test section has an opening, through which the top endplate is attached to the load balance flexure. The force balance uses a strain gauge bridge providing voltage outputs proportional to lift and drag forces. Lift forces are measured by four strain gauges (two on each side), and drag forces are measured by two strain gauges (one per side). The output from the load cells is sampled for a minimum of 30 s at each flow condition at a rate of 1.6 kHz to obtain stable force data. The measured voltage signals are transferred to a data acquisition computer through a 24-bit strain gauge module (NI-9237, National Instruments, Austin, TX) for post-processing. The measurements are repeated at least 10 times for each condition for statistical independence. The lift and drag coefficients are calculated as

$$C_L = \frac{L}{\frac{1}{2}\rho U_\infty^2 cb}, \quad (2)$$

$$C_D = \frac{D}{\frac{1}{2}\rho U_\infty^2 cb}, \tag{3}$$

where $L$ and $D$ are the time-averaged lift and drag, respectively, $\rho$ is the air density, and $b$ is the span of the airfoil model. The measurement range is 711 N for the lift force and 106 N for the drag force, with an accuracy of 0.04% and 0.03% of the full range, respectively. The resulting uncertainties in the force coefficients are less than 3%. The pitching moment coefficient is calculated using Eq.(4) about quarter chord location based on the measured lift force at the leading and trailing edges of the airfoil.

$$C_M = \frac{M}{\frac{1}{2}\rho U_\infty^2 cb}, \tag{4}$$

where $M$ is the pitching moment calculated at the quarter chord.

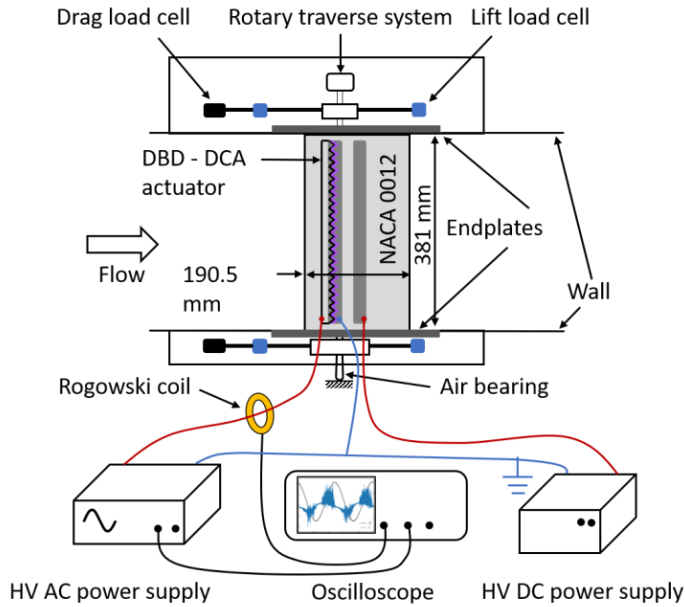

Figure 1. Schematic of the experimental setup for the lift, drag, and pitching moment measurements on the airfoil with sawtooth DBD – DCA actuator.

## 2.2. Plasma Actuator and Power Supply System

The plasma actuator consists of three 0.05 mm thick copper electrodes separated by four layers of Kapton film (7700 VPM @ 25 °C) of thickness 0.088 mm as shown in Figure 2(a). The exposed high voltage electrode is fabricated using electro-discharge machining to produce a sawtooth pattern, as shown in Figure 2 (b). The sawtooth pattern was selected based on a preliminary optimization study (Vaddi, Mamishev and Novosselov 2021). Gao et al. also reported that sawtooth a pitch-to-height ratio of 1 is favorable for power consumption, dielectric heating, and velocity induced by plasma (Guoqiang Gao 2017). The trough of the sawtooth is rounded to eliminate the high electric field concentration. The exposed and encapsulated electrodes are overlapped such that the ground electrode edge is aligned with the troughs of the sawtooth pattern. The width of the ground electrode is 25 mm and is long enough to allow for the development of the plasma sheet. The third electrode with a width of 25 mm, is placed 20 mm from the downstream edge of the ground electrode. The spanwise length of the electrodes is 350 mm. The actuator is installed at the 18% chord from the leading edge ($x/c = 0.18$) as shown in Figure 2 (c). The location is selected based on the reported position of separation bubble for NACA 0012 airfoil: $0.1\ c - 0.6\ c$ for $\alpha = 2^0 - 8^0$ at $Re = 3 \times 10^5$ (Winslow et al. 2018).

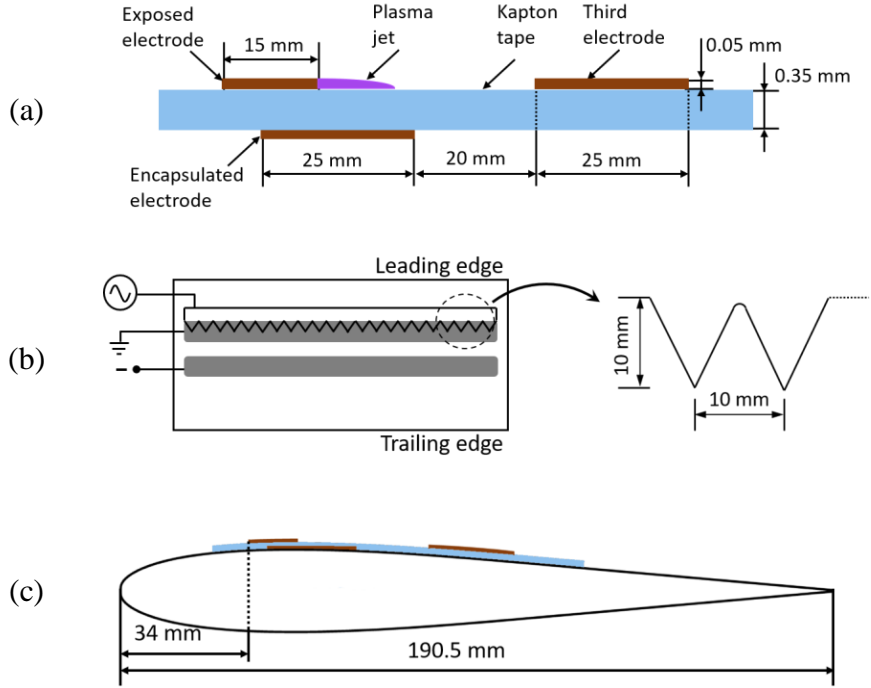

Figure 2. DBD – DCA actuator mounted on NACA 0012 airfoil (a) dimensions of the plasma actuator, (b) geometry of the sawtooth electrode pattern, and (c) actuator position on the airfoil.

The exposed electrode is connected to the HV AC power amplifier Trek 615-10 (Advanced Energy, Lockport, New York), see Figure 1. The input signal is the sinusoidal wave with the peak-to-peak voltage $\varphi_{p-p} = 18$ kV with a 1.25 kV DC bias voltage and frequency $f = 2$ kHz. The current is measured using Pearson model 8590C current probe positioned around the wire driving the exposed electrode. The current monitor is connected to a Tektronix DPO2024 oscilloscope that uses a bandwidth of 200 MHz to satisfy the Nyquist condition for achieving a sampling rate of 400 MS/s required for the accurate capture of individual discharges with a typical duration of ~ 30 ns (Tang, Vaddi, Mamishev and Novosselov 2021). The high bandwidth and the sampling rate minimize the noise during the current measurements and can be used to compute the time-averaged electrical power (Moreau 2007). The voltage from the power supply is also measured simultaneously. The third electrode is connected HV DC power supply (Bertan 205B-20R) with variable negative DC voltage $\varphi_{DC} = -(0-15)$ kV to the third electrode, and the current is measured directly from the power supply.

## 3. RESULTS AND DISCUSSION

### 3.1. Baseline Performance

The thin plasma actuator installed on the airfoil may affect the boundary layer development similar to a thin trip (Traub 2011); thus, the effect of a passive actuator (Plasma OFF) on the aerodynamic characteristics is explored first. These tests also provide validation of force measurements. The lift and drag coefficients vs. the angle of attack are plotted in Figure 3. The data is compared with the previous reports for 2D NACA0012 airfoil, e.g., data for $Re$ = 160k – 200k (Ladson 1988, Sheldahl and Klimas 1981) and numerical simulations (XFOIL) for $Re$ = 200k (Drela 1989). The lift coefficient at $Re$ = 201k shown in Figure 3(a) agrees well with the previous literature. The actuator in its OFF state does not change the aerodynamic characteristics of the airfoil. Other studies reported that thin strips (Traub 2011) and low profile vortex generators (Lin 2002) with a similar thickness to our actuator reported separation delay. Since we limit $\alpha < 8°$ the separation delay at higher angles has not been evaluated. The drag measurements in Figure 3(b) agree well with previous reports literature except for Ladson, which likely due to the difference in chord lengths. The chord length in Ladson data is 914 mm, while in our experiment, it is 190.5 mm. Our data, however, agrees well with the shorter chord length (150 mm) data of Sheldahl & Klimas.

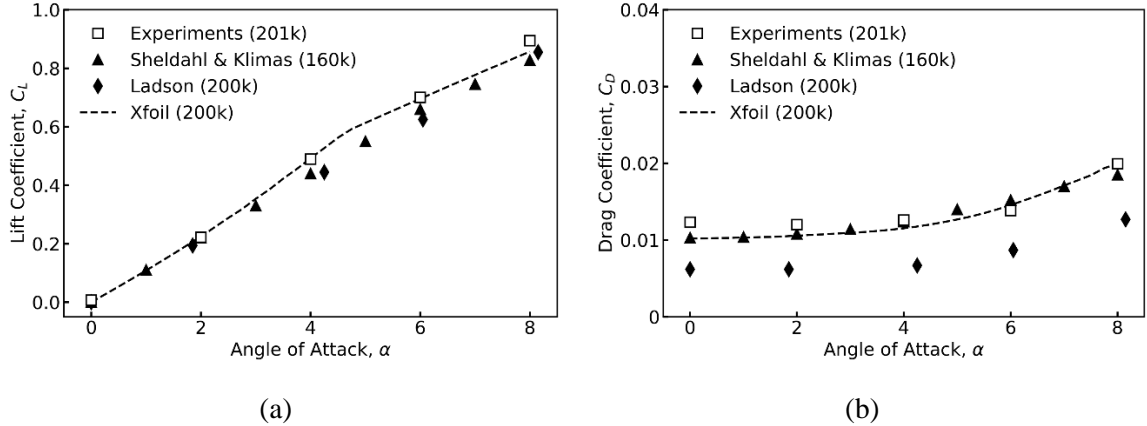

(a)            (b)

Figure 3. Comparison of measured aerodynamic coefficients for NACA 0012 airfoil, with XFOIL calculations (Drela 1989) and data from the literature (Ladson 1988, Sheldahl and Klimas 1981)

### 3.2. Effect of DCA Voltage

Vaddi et al. demonstrated the sawtooth DBD – DCA actuator mounted on a flat plate generates 4x more thrust than DBD by accelerating the ions in the negative DC field (Vaddi, Mamishev and Novosselov 2021). Greater momentum injected can be utilized in the airfoil case. Figure 4 shows the effect of the flow acceleration by DC field on aerodynamic coefficients for two different angles of attack at a free stream speed of 15 m/s. The voltage on the third electrode is varied from 0 kV to −15 kV. A plasma OFF case is given for reference. DBD is operated at $\varphi_{p-p}$ = 18 kV, $\varphi_{bias}$ = 1.25 kV and $f$ = 2 kHz. The bias voltage is applied to increase the amount of time with positive voltage during the AC cycle. All data is reported based on the instantaneous force measurement, i.e., within 3 seconds after the actuator is energized.

In all the cases, the lift coefficient increases with the DC field. Compared to the plasma-OFF condition, the sawtooth DBD – DCA with $\varphi_{DC}$= 0 kV (the third electrode grounded) case does not result in a significant lift, drag, or pitching moment change. As the magnitude of DCA voltage increases, significant changes are observed. The lift coefficient increases by ∼ 0.03 for both angles of attack at $\varphi_{DC}$ = −15 kV. The highest tested voltage case (−15 kV) corresponds to ∼7% increase in the lift at α = 4⁰. The lift augmentation can be explained by two mechanisms. (1) Momentum Injection: Stronger DC field accelerate the flow, similar to a corona discharge EHD flow (Guan et al. 2018, Vaddi, Guan, Mamishev and Novosselov 2020). The resulting jet propagates towards the trailing edge creating higher velocity flow on the suction side of the airfoil. (2) Vorticity Generation: Velocity is induced perpendicular to the sawtooth edge creates vortex structures as the oblique wall jets collide (Riherd and Roy 2013). The induced vortices trigger a rapid transition to the turbulent boundary and hinder the development of the laminar separation bubble. Vorticity and the momentum entertainment into the boundary layer locally accelerate the flow on the succession side of the airfoil, creating additional lift.

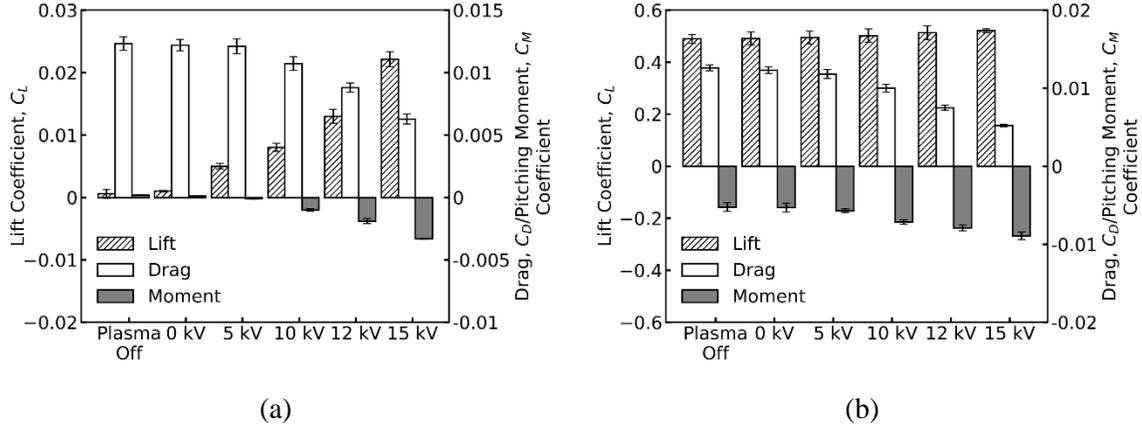

(a) (b)

Figure 4. Aerodynamic performance at different $\varphi_{DC}$ potentials for two different angles of attack (a) 0° and (b) 4° ($U_\infty$ = 15 m/s), DBD - DCA is operated at $\varphi_{p-p}$ = 18 kV, $\varphi_{bias}$ = 1.25 kV and $f$ = 2 kHz

The drag reduction on the airfoil, as seen in Figure 4 can be attributed to the momentum injection (i.e., thrust) and the vorticity generation by the sawtooth DBD - DCA actuator. Experiments with the straight edge electrode can potentially elucidate the individual contributions of each mechanism. For α = 0° case, the first significant change in $C_D$ (12% drop) is at $\varphi_{DC}$ = −10 kV. However, $C_D$ drops by 50% at $\varphi_{DC}$ = −15 kV. The total change from plasma OFF to maximum DBD – DCA input is $\delta C_D$ ~ 0.006. For α = 4° case, this difference is greater $\delta C_D$ ~ 0.0074, suggesting that DBD – DCA actuator alters the flow field on the suction side of the airfoil, such as removing the separation bubble leading to additional drag reduction.

One of the major findings is the effectiveness of the sawtooth DBD – DCA to alter the pitching moment. The momentum injection and vorticity generation at the trailing edge creates a greater pitching moment, which increases at the higher $\varphi_{DC}$ voltage as expected. The pitching moment changes by ~0.0035 for both angles of attack with no plasma actuation and plasma actuation with $\varphi_{DC}$ = −15 kV. The results show that the control efficiency has a strong dependence on DCA electrode potential. The position of the actuator will also affect the pitching moment; however, this topic is outside the current scope.

### 3.3. DBD-DCA: Function of Angle of Attack

To maximize the effect of the actuator on the airfoil performance, for the next set of experiments, the DCA voltage is held at −15 kV. The performance is assessed based on the relative difference between plasma ON and plasma OFF conditions. The results from the plasma actuation as a function of α are presented in Figure 5, with $U_\infty$ = 15 m/s. The experiments are conducted at $\varphi_{p-p}$ = 18 kV, $\varphi_{bias}$ = 1.25 kV and $f$ = 2 kHz. As a function of angle of attack, the lift increase is nearly constant, see Figure 5(a). At α = 0° with actuator ON, the improvement in lift coefficient is ~ 0.03. Similar lift trends have been reported with plasma gurney flaps (Feng, Shi and Liu 2017, Zhang, Liu and Wang 2009). However, the drag force increases when plasma gurney flaps are introduced. The sawtooth DBD – DCA actuator generates both significant lift and reduction in drag at low angles of attack. The lift increase $\delta C_L$ = 0.03 – 0.04 while reducing drag coefficient $\delta C_D$ ~ (0.006 – 0.008) over the range of α = 0° – 8°, see Figure 5(b).

The quarter chord pitching moment is shown in Figure 5(c). With the lift enhancement, the pitching moment has a negative value for all angles of attack tested. The pitching moment changes by 0.003 – 0.005 when the actuator is energized. The plasma ON condition for the DBD-DCA actuator mounted on the suction side of the airfoil leads to a greater negative pitching moment (nose-up) for all angles of attack.

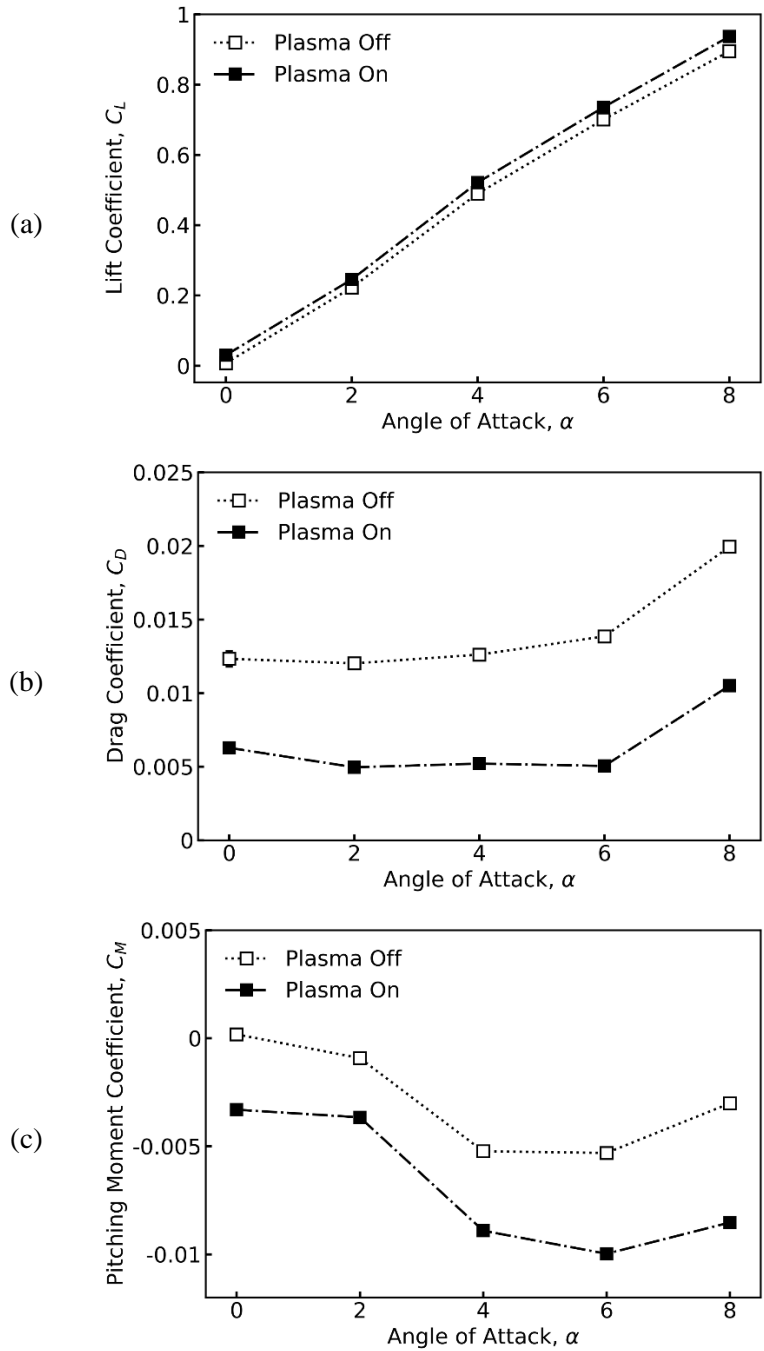

Figure 5. Aerodynamic characteristics (a) Lift (b) Drag (c) Pitching moment with/without plasma actuation at $U_\infty$ = 15 m/s. DBD - DCA is operated at $\varphi_{p-p}$ = 18 kV, $\varphi_{bias}$ = 1.25 kV, $\varphi_{DC}$ = −15 kV and $f$ = 2 kHz.

## 3.4. Effect of Reynolds Number

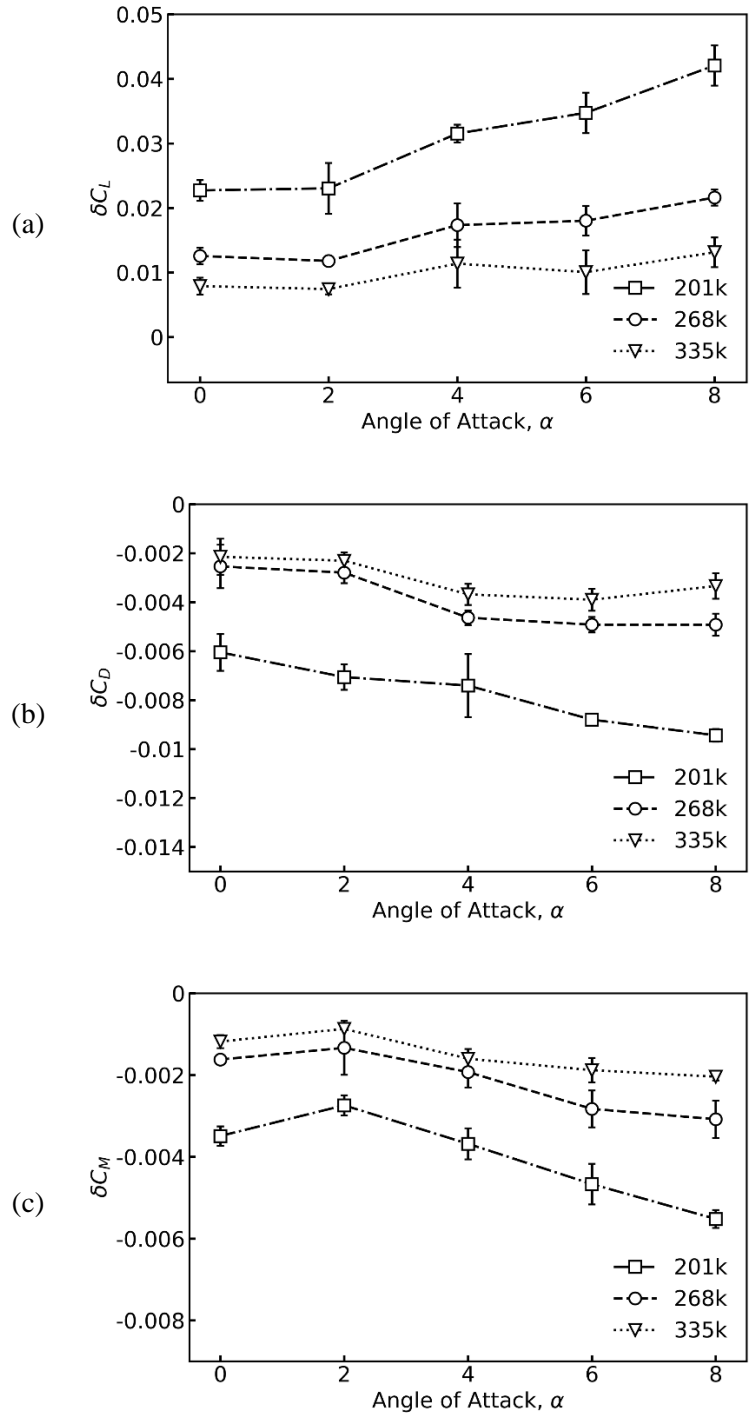

Figure 6. Effect of Reynolds number on (a) Lift (b) Drag (c) Pitching Moment for sawtooth DBD – DCA actuator.

The freestream velocities were varied from 15 m/s to 25 m/s, corresponding to the chord $Re$ = 201k – 335k. Figure 6 shows the effect of the Reynolds number on airfoil performance with sawtooth

DBD – DCA actuator. Figure 6(a) shows that the change in the lift ($\delta C_L$) for the lowest wind speed increases with α. However, DBD-DCA actuation at higher velocities $\delta C_L$ has only a slight increase with the angle of attack at a given Reynolds number. The effect of plasma actuation decreases with an increase with free stream velocity, which agrees with previous reports of maximum lift scaling as $\delta C_{L,max} \propto U_\infty^2$ (Corke et al. 2006). The change in drag coefficient as a function of $Re$ is shown in Figure 6(b). As the free stream velocity increases, the drag reduction ($\delta C_D$) decreases. At $U_\infty$ = 25 m/s, plasma ON condition results in a drag reduction of about 20%, compared to 50% at $U_\infty$ = 15 m/s. The $\delta C_D$ is also slightly decreases as a function of angle of attack for all wind speeds.

The pitching moment change ($\delta C_M$) is shown in Figure 6(c), for the actuator location considered in this work, with the Reynolds number is similar to $\delta C_D$. The $\delta C_M$ slightly decreases with the angle of attack for all wind speeds. The effectiveness of the actuator for control purposes diminishes at higher velocities; however, the optimization of actuator location can lead to additional improvement in $\delta C_M$. With proper aircraft design, this metric could provide sufficient control authority for smaller UAVs.

## 4. CONCLUSION

This study provides a proof-of-concept demonstration of a novel DBD – DCA actuator for active flow control utilizing NACA 0012 airfoil at subsonic flow speeds with $Re$ = 201k – 335k and low angles of attack. The actuator was mounted in co-flow orientation on the suction side of the airfoil: the active AC electrode was positioned at 18% chord and the DCA electrode at 48% chord. Effects of DCA potential, angle of attack, wind speeds on airfoil performance were investigated. The DCA potential increases the lift coefficient, reduced drag, and significantly affects the pitching moment of the airfoil. These effects can be explained by two mechanisms (1) increase of momentum injection due to DCA effect, (2) Vorticity generation due to a sawtooth-shaped electrode.

The lift curve with plasma actuation is almost parallel to no actuation for all angles of attack $\alpha$ = $0^0$ – $8^0$ and the change in $C_L$~ 0.03 – 0.04. The $\delta C_L$ with and without plasma actuation decreases with free stream velocity, and the scaling factor agrees with the literature. The drag coefficient decreases with the $\varphi_{DC}$ and a drag reduction of 50% is measured when $\varphi_{DC}$ = −15 kV at $U_\infty$ = 15 m/s. Consistent $\delta C_M$ shows DBD – DCA can be used for active flow at low angles of attack for low-velocity UAVs. One of the limitations of plasma actuators is their ineffectiveness at high velocities; future research should address the performance of DBD-DCA actuators at higher wind speeds. To improve control authority, the research should explore different positions and orientations of the actuator. Another topic of interest is gain insight into the interaction between the free flow and the momentum injection into the flow boundary layer for sawtooth DBD – DCA either by flow visualization or surface pressure measurements.

## ACKNOWLEDGMENTS

We wish to thank Professor James Riley for the valuable discussion. This work was supported through an academic-industry partnership between Aerojet Rocketdyne and the University of Washington funded by the Joint Center for Aerospace Technology Innovation (JCATI) and is also based upon work supported in part by the Office of the Director of National Intelligence (ODNI), Intelligence Advanced Research Projects Activity (IARPA), via ODNI Contract 2017-17073100004. The views and conclusions contained herein are those of the author and should not be interpreted as necessarily representing the official policies or endorsements, either expressed or implied, of ODNI, IARPA, or the U.S. Government.